\documentstyle[12pt,epsf]{article}

\begin{document}

\begin{titlepage}

\begin{flushright}
KUNS-1370\\
HE(TH)95/20\\
hep-ph/9511395
\end{flushright}

\vskip 0.35cm
\begin{center}
{\large \bf
Duality of a Supersymmetric Standard Model without R parity
}
\vskip 1.2cm
Nobuhiro Maekawa
\footnote
{e-mail: maekawa@gauge.scphys.kyoto-u.ac.jp}
and Joe Sato
\footnote
{e-mail: joe@gauge.scphys.kyoto-u.ac.jp}
\vskip 0.4cm

{\it Department of Physics, Kyoto University,\\
      Kyoto 606-01, Japan}

\vskip 1.5cm

\abstract{
Recently one of the authors proposed a dual theory of a
Supersymmetric Standard Model (SSM), in which it is naturally
understood that at least one quark (the top quark) should be heavy, i.e.,
almost the same
order as the weak scale, and
the supersymmetric Higgs mass parameter $\mu$
can naturally be expected to be small.
Unfortunately, the model cannot possess Yukawa couplings of lepton sector.
In this paper,
we examine a dual theory of a Supersymmetric Standard Model without R
parity.
In this scenario,  we can introduce Yukawa
couplings of lepton sector. In order to induce the enough large Yukawa
couplings of leptons, we must introduce fairly large R parity breaking
terms, which may be observed in the near future.
}

\end{center}
\end{titlepage}

%
%
%
%

Recently, it has become clear that certain aspects of four dimensional
supersymmetric field theories can be analyzed exactly
\cite{duality,holomorphy,both,DSB}. By using the innovation, it has
been tried to build models in order to solve some phenomenological
problems\cite{DSB,mike,IzaYana,Stras}.
One of the most interesting aspects is ``duality''
\cite{duality,both}.
By using
``duality'', we can infer the low energy effective theory of a strong
coupling gauge theory.
One of the authors suggested that nature may use this ``duality''.
He discussed a duality of a Supersymmetric Standard
Model(SSM). Unfortunately, his model does not possess Yukawa couplings
of lepton sector. One possibility to introduce
them is to unify quarks and leptons by
considering the Pati-Salam gauge group\cite{MT}
In this paper, we discuss the model
with R-parity brealing terms in order to
obtain the Yukawa couplings of the lepton sector.

First we recaptulate the previous model. Then
we discuss the extention of the previous model and
see how leptons acquire their mass. Then we give
a summary and discussion.

To get the point of the previous idea
first we review Seiberg's duality.
Following his discussion
\cite{duality}, we examine $SU(N_C)$ Supersymmetric(SUSY)
QCD with $N_F$ flavors of chiral superfields,
\vskip 0.5cm
\begin{center}
\begin{tabular}{|l|c|c|c|c|c|} \hline \hline
 & $SU(N_C)$ & $SU(N_F)_L$ & $SU(N_F)_R$ & $U(1)_B$ & $U(1)_R$ \\ \hline
$Q^i$    & $N_C$ & $ N_F$ &  1  &   1    & $(N_F-N_C)/N_F$  \\
$\bar Q_j$ & $\bar N_C$ & 1 & $\bar N_F$ & $-1$ & $(N_F-N_C)/N_F$ \\
\hline \hline
\end{tabular}
\end{center}
\vskip 0.5cm
which has the global symmetry $SU(N_F)_L\times SU(N_F)_R \times U(1)_B
\times U(1)_R$.
In the case $N_F\geq N_C+2$, Seiberg suggested
\cite{duality} that at the low energy scale the above theory is
equivalent to the following $SU(\tilde N_C)$ SUSY QCD theory $(\tilde
N_C=N_F-N_C)$ with $N_F$ flavors of chiral superfields $q_i$ and $\bar
q^j$ and meson fields $T^i_j$,
\begin{center}
\begin{tabular}{|l|c|c|c|c|c|} \hline \hline
       & $SU(\tilde N_C)$ & $SU(N_F)_L$ & $SU(N_F)_R$ & $U(1)_B$ &
       $U(1)_R$ \\ \hline
$q_i$    & $\tilde N_C$ & $ \bar N_F$ &  1  &  $N_C/(N_F-N_C)$ & $N_C/N_F$  \\
$\bar q^j$ & $\bar {\tilde N_C}$ & 1 & $ N_F$ & $-N_C/(N_F-N_C)$ & $N_C/N_F$ \\
$T^i_j$ & 1 & $N_F$ & $\bar N_F$ & 0 & $2(N_F-N_C)/N_F$ \\
\hline \hline
\end{tabular}
\end{center}
\vskip 0.5cm
\noindent
and with a superpotential
\begin{eqnarray}
W=q_iT^i_j\bar q^j.
\end{eqnarray}

Then the idea of SSM is the following
\cite{mike}.

We introduce ordinary matter superfields
besides Higgs doublets:
\begin{eqnarray}
&&Q^i_L=(U_L^i, D_L^i):(3,2)_{1\over 6},\quad U_{Ri}^{c}:(\bar
3,1)_{-{2\over 3}},\quad
D_{iR}^{c}:(\bar 3,1)_{1\over 3} \nonumber \\
&&L^i=(N_L^i,E_L^i):(1,2)_{-{1\over 2}},\quad
E_{Ri}^{c}:(1,1)_1,\quad\quad i=1,2,3,
\end{eqnarray}
which transform under the gauge group $SU(3)_{\tilde C}\times
SU(2)_L\times U(1)_Y$.

Let's examine the dual theory of this theory with respect
to the gauge group
$SU(3)_{\tilde C}$. In this case the number of the flavor
is 6 and the global symmetry in the sense of Seiburg's is
$SU(6)_{QL}\times SU(6)_{QR}\times U(1)_B\times U(1)_R$.
We can assign $Q=(U_L^1,D_L^1,U_L^2,D_L^2,U_L^3,D_L^3)$ and $\bar
Q=(U_R^{c1},D_R^{c1},U_R^{c2},D_R^{c2},U_R^{c3},D_R^{c3})$
in the table.
Since $N_F=6$, the dual gauge
group is also $SU(3)_C$ ( $\tilde N_C=N_F-N_C$ ), which we
will assign to the ordinary QCD gauge group.
A subgroup, $SU(2)_L\times U(1)_Y$, of
the global symmetry group $SU(6)_{QL}\times
SU(6)_{QR}\times U(1)_B\times U(1)_R\times SU(6)_{L}\times
SU(3)_{ER}\times U(1)_L\times U(1)_{ER}$ is  gauged.
$ SU(6)_{L}\times
SU(3)_{ER}\times U(1)_L\times U(1)_{ER}$ is the global
symmetry of the lepton sector. For example, $ SU(6)_{L}\times U(1)_L$
acts on the multiplet $L \equiv (N^1_L, E^1_L,N^2_L, E^2_L,N^3_L,
E^3_L)$.
$SU(2)_L$ ge
nerators are given by
\begin{eqnarray}
I_L^a=I_{QL1}^a+I_{QL2}^a+I_{QL3}^a
+I_{L1}^a+I_{L2}^a+I_{L3}^a,\quad a=1,2,3,
\end{eqnarray}
where $I_{QLi}^a$ [$I_{Li}^a$]
are generators of $SU(2)_{QLi}$ [$SU(2)_{Li}$] symmetries which
rotate $(U_L^i, D_L^i)$ $[(N^i_L, E^i_L)]$.
and the generator of hypercharge $Y$ is given by
\begin{eqnarray}
Y={1\over 6}B-(I_{R1}^3+I_{R2}^3+I_{R3}^3)
-{1\over 2}L +ER
\end{eqnarray}
where $I_{QRi}^a$ are generators of $SU(2)_{Ri}$ symmetries which
rotate $(U_{Ri}^c, D_{Ri}^c)$.
In this theory, the global symmetry group is $SU(3)_{QL}\times
SU(3)_{UR}\times SU(3)_{DR}\times U(1)_B\times U(1)_R
\times SU(3)_{L}\times
SU(3)_{ER}\times U(1)_L\times U(1)_{ER}$
\footnote{Strictly speaking two of four global $U(1)$'s are broken
by $SU(2)_L$ and $U(1)_Y$ anomalies.}.
Then we can write down the quantum numbers of dual fields;
\begin{eqnarray}
&&q_{Li}=(d_{Li}, -u_{Li}):(3, \bar 2)_{1\over 6},\quad u_R^{ci}:
(\bar3,1)_{-{2\over 3}},\quad
d_R^{ci}:(\bar 3,1)_{1\over 3} \nonumber \\
&&M^i_j:(1,2)_{-{1\over 2}},\quad N^i_j:(1,2)_{1\over 2}\\
&&L^i=(N_L^i,E_L^i):(1,2)_{-{1\over 2}},\quad
E_{Ri}^{c}:(1,1)_1,\quad\quad i=1,2,3,
\nonumber
\end{eqnarray}
under the standard gauge group $SU(3)_C\times SU(2)_L\times U(1)_Y$.
Here $M^i_j\sim Q_L^iU_{Rj}^c$ and $N^i_j\sim Q_L^iD_{Rj}^c$ are the
meson fields and we assign
$q=(d_L^1, -u_L^1, d_L^2, -u_L^2, d_L^3, -u_L^3)$ and $\bar
q=(d_{R}^{c1}, -u_{R}^{c1}, d_{R}^{c2}, -u_{R}^{c2}, d_{R}^{c3}, -u_R^{c3})$.
Because leptons don't have color indices they exist as they were.
It is interesting that the matter contents of both theories are almost
the same. The difference is the existence of nine pairs of Higgs
superfields $M^i_j$ and $N^i_j$
and their Yukawa terms coupling to ordinary matter superfields,
\begin{eqnarray}
W=-q_L^i N_i^j u_{Rj}^c+q_L^iM_i^jd_{Rj}^c.
\end{eqnarray}

If one combination of these Higgs scalar fields
\footnote{In this paper we denote scalar component
by tilde, $e.x$ $\tilde A$ means the scalar component of
the superfield $A$.}
\begin{eqnarray}
\tilde H=\sum_{i,j}^3 a^i_j \tilde N_i^j+b^i_j (\tilde M^c)_i^j,
\end{eqnarray}
where $M^c$ denotes the charge conjugated field of $M$,
has  a VEV $\langle H\rangle=(v,0)$, the $SU(2)_L\times U(1)_Y$
symmetry is broken to
the electromagnetic gauge group $U(1)_Q$.
In this case, ordinary quark mass matrices are determined by the mixing
of the Higgs scalar field.
You should notice that at least one quark has a heavy
mass, which is almost the order of the weak scale $v$, if the Yukawa
coupling can be taken to be of order one because of the strong
dynamics.  Namely
the heaviness of the top quark can be naturally
understood.

There is, however,
a serious problem that the leptons are massless in
this theory.

We break R parity in the above model
in order to induce the Yukawa couplings of leptons.
We introduce the superpotential
\begin{eqnarray}
W=\lambda_{ijk}L^iL^jE_R^{ck},
\end{eqnarray}
which breaks the global symmetry $SU(3)_L\times SU(3)_{ER}\times
U(1)_L\times U(1)_{ER}$ for leptons and R parity.
Because leptons are singlet under the $SU(3)_{\tilde C}$
these terms survive in the dual theory.
Though to introduce
all of such terms causes dangerous phenomena which
are experimentally excluded, we can introduce
only one term among them\cite{DHBGH,HS}.

In the theory with $SU(3)_{\tilde C}$, we introduce the following
R-violating superpotential,
\begin{eqnarray}
W=\lambda L^2L^3E_{R}^{c3},
\label{eq:LBreaking}
\end{eqnarray}
which breaks the global symmetry of the lepton sector $SU(3)_L\times
SU(3)_{ER}\times U(1)_L\times U(1)_E$ to $SU(2)_{L23}
\times U(1)_{L1} \times SU(2)_{ER} \times U(1)_{(E1+E2)}
\times U(1)_{(L2+L3-2E3)}$, where $SU(2)_{L23}$ acts on the multiplet
$(L_2,L_3)$ , $U(1)_{L1}$ means that the charge of $L_1$ is 1 and
those of others' are 0 and so on.
The magnitude of the coupling is expected to be O(0.1)\cite{DHBGH}.

If the scalar component of
$L_i$'s ($\equiv \tilde{L_i}$)
get a VEV, then at least one lepton acquire mass.
When we consider SUSY breaking terms in the theory
it is naturally understood that $L_i$'s acquire
VEVs. We will see this in the following.

First of all, the global symmetry group
$SU(3)_{QL}\times SU(3)_{UR}\times SU(3)_{DR}\times U(1)_R$
must be broken explicitly because otherwise
when Higgs fields $N_i^j$ and $M_i^j$ have VEVs,
that is the global symmetry is
spontaneously broken,  massless Nambu-Goldstone bosons
appear.
One possibility of breaking the global symmetry explicitly
is to introduce soft SUSY breaking terms\cite{Evans,Peskin}
which also break
all the global symmetry except $U(1)_B$
\footnote{We introduce SUSY breaking terms which do not
respect the holomorphy. Such a term does not cause
quadratic divergence, that is, does not spoil
the hierachy stability
unless there is a singlet field. We assume simply
that there is no singlet field.}:
\begin{eqnarray}
\sum_{i,j,k} ( A_{ijk} \tilde Q_i \tilde U^c_j \tilde L_k
+  B_{ijk} \tilde Q_i \tilde D^c_j \tilde L_k^*
+  C_{ijk} \tilde L_i  \tilde L_j \tilde E_k)
+ \hbox{mass term}.
\end{eqnarray}

Thus there appear mixing terms between
the Higgs doublets, $\tilde N_i^j$ and $\tilde M_i^j$, and
the scalar components of lepton doublets $\tilde{L_i}$:
\begin{eqnarray}
m^2_{N_i^jk} \tilde N_i^j \tilde{L_k}
+m^2_{M_i^jk} \tilde M_i^{j*} \tilde{L_k},
\end{eqnarray}
as well as self-mass of $\tilde{L_i}$
\begin{eqnarray}
  m^2_{\tilde{l}i } |\tilde{L_i}|^2.
\end{eqnarray}

Then after Higgs fields get VEVs, VEVs of the scaler leptons
are induced. We assume transition masses are smaller
than self masses because the former breaks not only
the global flavor symmetry but also the lepton global
symmetry while the latter breaks only some part of
the lepton global symmetry.
In this case the induced VEVs are roughly given by
$<\tilde{L_i}> \sim (m^2_{N_j^ki} / m^2_{\tilde{l}i }) <N_j^k> \sim
(m^2_N / m^2_{\tilde{l}i }) v$, where $m^2_N$
is a typical transition mass and $v$ is the VEV of $\tilde H$.
Graphically they are expressed in Fig. 1(a).
Naively the magnitude of the scalar lepton VEVs is O(10) GeV.

\begin{figure}
\unitlength=1cm
\begin{picture}(16,5)
\unitlength=1mm
\put(4,15){$\tilde L_i$}
\put(22,15){$m^2_{N_i^jk} (m^2_{M_i^jk})$}
\put(54,15){$\tilde N_i^j ( \tilde M_i^{j*}) $}
\put(86,15){$L_3 (L_2)$}
\put(122,15){$\lambda$}
\put(156,15){$E^{c3} $}
\put(116,24){$\tilde L_2$}
\put(126,24){$(\tilde L_3) $}
\centerline{
\epsfxsize=16cm
\epsfbox{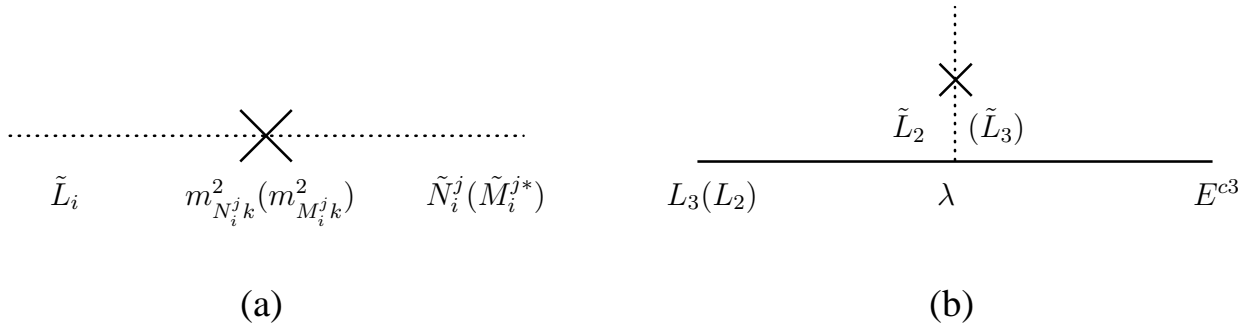}
}
\end{picture}
\caption{(a) a typical graph which gives a VEV to sneutrinos (neutral
component of $\tilde L_i$). (b) a graph which induces a mass term
for lepton at tree level below the electroweak scale.
}
\end{figure}

By these VEVs lepton masses arise at tree level
according to (\ref{eq:LBreaking}) (see Fig 1(b)).
The form of the mass matrix for leptons is
\begin{eqnarray}
M_l=\pmatrix{0&0&0\cr
0&0&m_2\cr
0&0&m_3}
\label{LeptonMass}
\end{eqnarray}
where $m_2 = - \lambda <\tilde{L_3}>, m_3 = \lambda <\tilde{L_2}>
\sim$ O(1) GeV\footnote{Because there is $SU(2)_{L23}$ global
symmetry, we can assume, without loss of generality,
that among $\tilde L_2$ and  $\tilde L_3$
only $\tilde L_2$ acquires a VEV and hence
only $m_3$ is not zero. }.
Thus we can understand
$\tau$ mass naturally.

\begin{figure}
\unitlength=1cm
\begin{picture}(16,7)
\unitlength=1mm
\put(40,12){$ L_i$}
\put(62,10){$\tilde \gamma$}
\put(77,10){$m_{\tilde \gamma}$}
\put(90,10){$\tilde \gamma$}
\put(114,12){$E^{cj} $}
\put(52,36){$\tilde L_i$}
\put(76,60){$<\tilde L_l>$}
\put(98,36){$\tilde E_j$}
\put(80,45){$C_{ilj}$}
\centerline{
\epsfxsize=8cm
\epsfbox{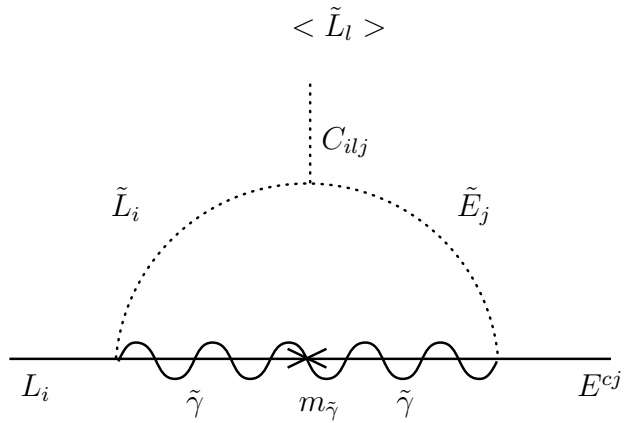}
}
\end{picture}
\caption{a typical graph which  induces a mass term
for lepton at ine loop below the electroweak scale.
}
\end{figure}

Through radiative corrections like Fig. 2 \cite{BHRS}
the other will be induced. These are quite dependent on
$C_{ijk}$\footnote{In the case that R-Parity is broken,
in general, neutrino masses are also induced\cite{HS}.
These are also dependent on SUSY breaking parameters
and here we do not
touch the detail.}.

In summary, we examine duality of a SUSY model with
R-Parity breaking terms. In this model,
the Yukawa couplings of the lepton sector as well as
the quark sector are indeuced. Moreover since
all the global symmetries except $U(1)_B$ are broken
by SUSY breaking terms, we can avoid the appearance of Nambu-Goldstone
bosons when Higgs fields aquire VEVs. The fact that the nature
doesn't respect the R-Parity, which might be observed in near future,
may suggest that the nature uses the duality.

\section*{Acknowledgements}

We are grateful to the organizers of the 1995 Ontake Summer Institute
and to M. Peskin for a stimulating set of lectures at the institute.
We would like to thank our colleagues for
discussions on ``duality''.
N.M. also thanks T.~Kawano and M.Strassler for useful discussions.
J.S thanks K.~Inoue for valuable discussion.

\end{document}